# Quantum vs. Classical Spin: A Comparative Study of Dipolar Spin Dynamics and the Onset of Chaos

Victor Henner[1,2], Alexander Nepomnyashchy[3], Tatyana Belozerova[2]

[1]Department of Physics, Loyola University, Chicago IL, 60660, USA

[2]Department of Physics, Perm State University, Perm, 614990, Russia

[3]Department of Mathematics, Technion –Israel Institute of Technology, 32000 Haifa, Israel[1]

**Abstract**

We investigate the spin dynamics of a dipole-coupled system by comparing a direct solution of the Schrödinger equation for quantum spins with simulations of classical spins. Although classical spins have long been used in microscopic spin dynamics simulations, we demonstrate that their results differ significantly from those of quantum spins. Using Free Induction Decay (FID) as a benchmark, we find that while the overall patterns are qualitatively similar, significant discrepancies emerge at both short and long timescales. We trace these differences to fundamental distinctions in the two descriptions. On short timescales, the dynamics of classical spins – governed by a continuous frequency spectrum that depends on initial conditions – are far more sensitive to initial configurations than those of the quantum system, which possesses a fixed, discrete spectrum. On long timescales, we demonstrate that the nonlinear classical system exhibits chaotic dynamics, as quantified by a positive Lyapunov exponent, which underlies its long-time divergence from quantum behavior.

## 1. Introduction

The correspondence between quantum and classical approaches to spin dynamics is a topic of fundamental importance in condensed matter physics, magnetism, and quantum information. Classical spin models are frequently employed to study dynamic phenomena such as magnetic resonance, free-induction decay (FID), Rabi oscillations, and relaxation. Although quantum mechanics provides the exact framework for describing spin systems, quantum simulations of spin dynamics quickly become computationally intractable as the number of spins increases. In contrast, the classical equations of motion can be numerically

---

[1] Email addresses: vhenner@luc.edu (V.Henner), nepom@technion.ac.il (A.Nepomnyashchy), tsbelozerova@yandex.ru (T.Belozerova)

integrated for systems of hundreds or even thousands of spins, making classical simulations a widely used tool in spin physics. The application of this approach to magnetic resonance dates to Jensen and Platz (1973) [1], who computed FID line shapes in dipolar-coupled rigid lattices using direct classical-spin simulations. Subsequent works have further investigated the dynamics of classical spins [2–4]. For instance, a study of classical and quantum spin chains with nearest-neighbor interactions demonstrated that classical simulations can, in some cases, yield quantitatively accurate results for FID [5]. In work [6], it was shown that the equations of motion for dipole-interacting spins can be mapped from a quantum to a classical form. Conversely, a method for quantum mechanical simulation based on the direct solution of the Schrödinger equation for spin-½ systems was suggested in [7], which also took initial steps toward comparing quantum and classical FID. Work [8] compared FID simulations based on spin-wave theory with direct classical-spin simulations. Furthermore, classical equations of motion have been used to study collective relaxation phenomena enhanced by radiation damping [9-12]. Collectively, these studies confirm that classical spins, while an approximation, capture many essential features of spin dynamics.

Nevertheless, significant differences persist between the results obtained from classical and quantum spin models. Understanding the origin of these discrepancies is the central subject of this paper. For this purpose, we focus on FID – a process of fundamental importance for both theory and applications. Our analysis is not tied to a specific spin system or experimental setting; rather, we aim to demonstrate the relativity of the agreements between quantum and classical descriptions, as well as the fundamental origins of their differences.

The underlying reason for these differences is that a quantum system has a fixed, discrete energy spectrum, whereas a classical spin system possesses a continuous spectrum that depends on the initial conditions.

Although classical spins have long been used in microscopic spin dynamics simulations, we demonstrate that their results differ significantly from those of quantum spins. To analyze the discrepancies at times longer than the characteristic dipole-dipole time, we examine chaotic behavior in a system of classical spins by evaluating the Lyapunov exponent. The issue of chaotic dynamics in spin systems has been explored in several works; for example, Reference [13] investigates the onset of chaos in ensembles of classical spins with nearest-neighbor couplings to determine the genericity of this behavior.

## 2. Quantum simulations

### 2.1 The quantum spins approach

For a system of $N$ spins all coupled by dipolar interactions, the quantum Hamiltonian for FID problem consists of the Zeeman and the dipolar parts:

$$\hat{\mathcal{H}} = -\gamma \hbar H_0 \sum_{j=1}^{N} \hat{S}_j^z + (\gamma \hbar)^2 \sum_{j<k} \frac{\hat{\vec{S}}_j \cdot \hat{\vec{S}}_k}{r_{jk}^3} - 3 \frac{(\hat{\vec{S}}_j \cdot \vec{r}_{jk})(\hat{\vec{S}}_k \cdot \vec{r}_{jk})}{r_{jk}^5}, \quad (1)$$

where $H_0$ is large static magnetic field along the $z$-axis, $\gamma = -g_e |e|/2m_e c$, $g_e = 2$. The magnetic moment operator is $\hat{\vec{\mu}} = -\gamma \hat{\vec{S}}$. The $j$-th spin is represented by spin-1/2 operators $\hat{\vec{S}}_j = (\hat{S}_j^x, \hat{S}_j^y, \hat{S}_j^z)$ with eigenvalues $\pm 1/2$. The vector $\vec{r}_{jk}$ connects the positions of spins $j$ and $k$.

The Hamiltonian (1) is time independent. We can write it as $\hat{\mathcal{H}} = \hat{\mathcal{H}}_Z + p_d \hat{\mathcal{H}}_d$, $\hat{\mathcal{H}}_Z = -H_0 \hat{S}^z$ with the parameter $p_d = \frac{\gamma \hbar}{a^3 H_0} = \frac{\mu}{a^3 H_0}$ ($a$ is the distance between neighboring spins, it is the lattice step in case of a regular lattice) representing the scale of the dipole energy in units of the Zeeman energy. This parameter can be also presented as $p_d = J/\omega_0$ with $J = \frac{\gamma^2 \hbar}{a^3}$ and $\omega_0 = \gamma H_0$.

The mathematical approach [7] we employ is exact and based on the first-quantum mechanical principles.

First, we find the eigenvectors and eigenvalues, $v_i$ and $E_i$, of operator (1):

$$\hat{\mathcal{H}} v = E v. \quad (2)$$

By diagonalizing the $\hat{\mathcal{H}}$ matrix and using $\Psi_i(t) = v_i \exp(-iE_i t)$, we construct the general solution for $N$ spins ½ (yielding $R = 2^N$ states):

$$\Psi(t) = \sum_{i=1}^{R} C_i \Psi_i(t) \quad (3)$$

of the time-dependent Schrödinger equation:

$$i \frac{d}{dt} \Psi(t) = \hat{\mathcal{H}} \Psi(t). \quad (4)$$

The constants $C_i$ are determined by the initial condition. If initially all spins are aligned along the $x$-axis, the initial wave function $\Psi_0$ is given by the equation

$$\hat{S}^x \Psi_0 = \frac{N}{2}\Psi_0, \qquad (5)$$

where $\hat{S}^x = \sum_i^N S_i^x$ with $S_i^x = I \otimes \ldots \otimes I \otimes S^x \otimes I \otimes \ldots \otimes I$; matrix $S^x = \frac{1}{2}\begin{pmatrix} 0 & 1 \\ 1 & 0 \end{pmatrix}$ is placed in the $i$-th position, $I = \begin{pmatrix} 1 & 0 \\ 0 & 1 \end{pmatrix}$.

After finding $\Psi_0$, we impose the initial condition:

$$\Psi(0) = \Psi_0, \qquad (6)$$

i.e.,

$$\sum_{i=1}^R C_i v_i = \Psi_0. \qquad (7)$$

Projecting both sides of (7) on $v_i$, we find $C_i$:

$$C_i = \langle v_i, \Psi_0 \rangle. \qquad (8)$$

Substituting the obtained values of $C_i$ into (3), we obtain the particular solution of the Schrodinger equation (4) satisfying the initial condition (5):

$$\Psi(t) = \sum_{i=1}^R C_i v_i \exp(-iE_i t). \qquad (9)$$

Now we can find the expectation value $S^x(t)$ as

$$S^x(t) = \langle \Psi(t), \hat{S}^x \Psi(t) \rangle = \Psi^*(t)\hat{S}^x \Psi(t) \qquad (10)$$

as well as

$$S^{y,z}(t) = \langle \Psi(t), \hat{S}^{y,z}\Psi(t) \rangle = \Psi^*(t)\hat{S}^{y,z}\Psi(t). \qquad (11)$$

These functions give the dynamics of the spin system.

### 2.2 Free induction decay

Free induction decay occurs because spins that are initially in phase gradually lose phase coherence due to spin–spin interactions. This dephasing leads to a completely random phase distribution in the transverse plane. As a result, the net transverse magnetization – the vector sum of all spins – goes to zero, even though individual spins continue to precess. This explains why the expectation value (the observable) in equation (10) decays, even though the time evolution of each individual spin wave function in equation (9) remains periodic.

A remarkable feature of the FID is that the initial transverse polarization oscillates with a decaying amplitude. The time $t_*$ at which the transverse polarization first reaches zero can be associated with the phenomenological dipole time, $T_2$.

Performing calculations with different values of $p_d$ we found that $t_*$ is practically inversely proportional to $p_d$. Just to specify, we use the value $p_d = 0.01$ to present the results in this section. All calculations are performed over a time interval much shorter than the spin-lattice relaxation time, so this interaction can be neglected.

The figures in this section show the transverse polarization evolution $S^x(t) = (1/N)\sum_l S_l^x(t)$ for a system of quantum spins sitting in the nodes of a regular lattice.

The function $S^x(t)$ is proportional to the free precession signal amplitude. Its Fourier transform, $S^x(\omega)$, gives the shape function usually denoted as $f(\omega)$, meaning that experimental measurements of FID signal and $f(\omega)$ are complementary.

The discrete Fourier transform (DFT) we use is defined as:

$$X_k = \sum_{n=0}^{M-1} x_n e^{-i\frac{2\pi k}{M}n}, \quad k = 0,1,...,M-1.$$

The inverse DFT is

$$x_n = \frac{1}{M}\sum_{k=0}^{M-1} X_k e^{i\frac{2\pi k}{M}n}, \quad n = 0,1,...,M-1.$$

Here $x_n$ is the value of $x(t)$ ($S^x(t)$ in our case) for $t = t_n$; $X_k = X(\omega_k = 2\pi k)$, and $M$ is the number of points (for $\tilde{t}_{max} = 2000$, as in Figs. 1 and 8, $M = 4000$).

The $f(\omega)$ is computed with a frequency resolution step $4.5 \cdot 10^{-4} \omega_0$. The spectrum is centered at $\omega_0$ with a width determined by the value of parameter $p_d$. On Figure 1 the FID calculation with 12 quantum spins and its FT are presented.

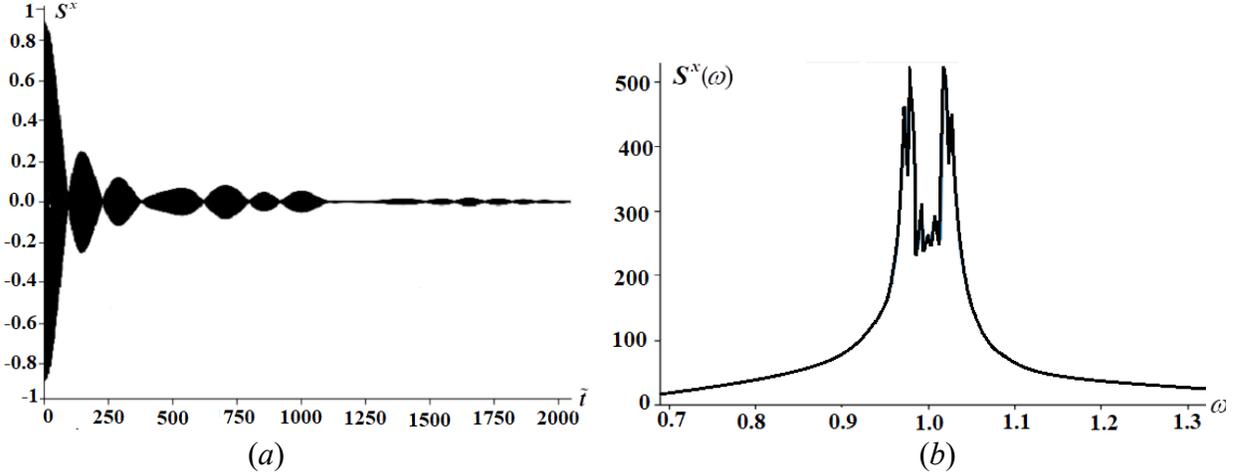

(a)                                                          (b)

Fig. 1. Results of simulations of FID with 12 quantum spins. Left panel $S^x(t)$ (normalized by unity), right panel – the modulus of the Fourier transform, $S^x(\omega)$ in arbitrary units. $p_d = 0.01$. The dimensionless time is $\tilde{t} = \omega_0 t$, $\omega_0 = \gamma H_0$ (in dimensionless units $\tilde{\omega}_0 = 1$).

The width of the resonance curve is about $p_d$, two-peaks symmetrically disposed about the Larmor frequency $\omega_0$ is the Pake doublet [14] - the result of altering the field $H_0$ by spin's nearest neighbors.

One can see that the FID amplitude convergence, $S^x(t) \to 0$, is very good. On a time interval of several $t_*$, $S^x(t)$ fits well to Abragam's trial function [15]

$$f(t) = A\exp\left(-\frac{a^2 t^2}{2}\right)\frac{\sin bt}{bt}. \tag{12}$$

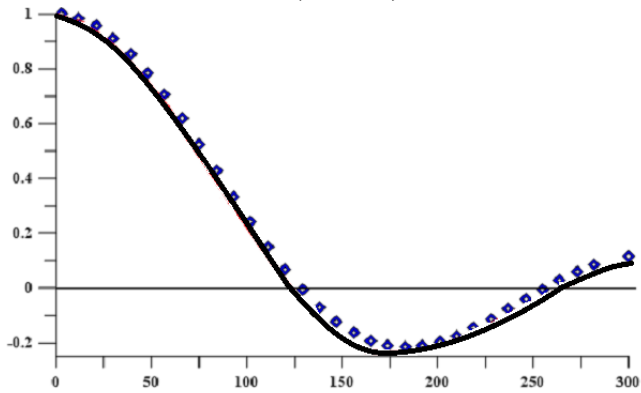

Fig 2. The envelope of $e^x(t)$ in Figure 1 (solid line) and Abragam's trial function $f(t)$ (squares) with parameters $A = 1$, $a = 0.00065$, $b = 0.0245$.

It is important to analyze the dependence of FID behavior on the number of spins. Notice that on a standard desktop computer no more than thirteen quantum spins ½ can be managed because of computer memory limitation. The corresponding graphs are presented in Figure 3.

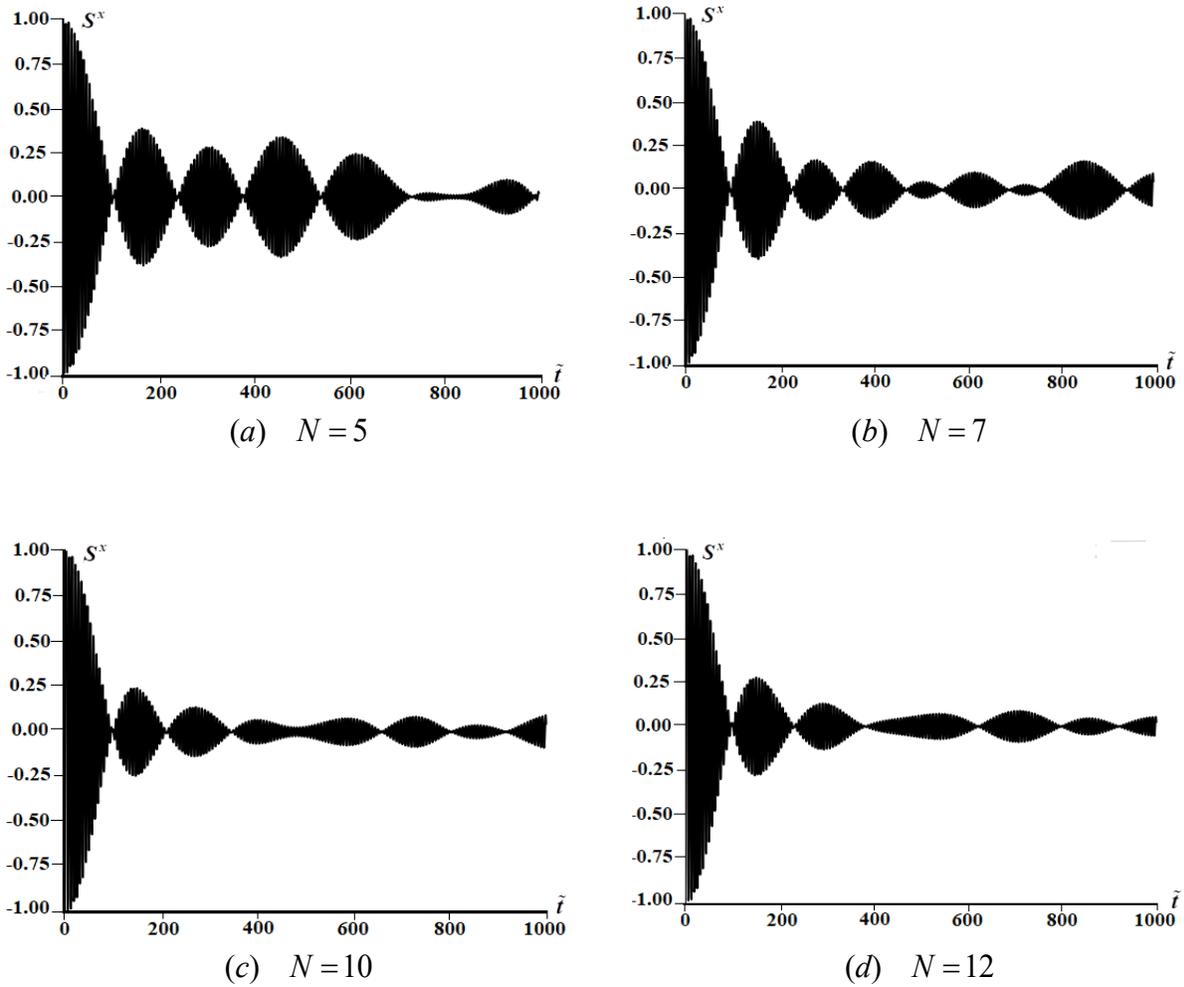

Fig.3. panel a: $N = 5$, spins Cartesian coordinates are [0,0,0], [0,0,1], [0,1,0], [1,0,0], [0,1,1].
Panel b: $N = 7$. Two more spins added at [1,0,1], [1,1,0].
Panel c: $N = 10$. Three more spins added at [1,1,1], [0,0,-1], [0,-1,0].
Panel d: $N = 12$. Two more spins added at [-1,0,0], [-1,0,-1].

For five spins there is already some collective decay of the initial amplitude, for seven spins a collective decay is getting more observed. Remarkably, that already for 10 spins the FID demonstrates satisfactory collective behavior, for 12 and 10 spins the results are close. Also, for $N = 12$ and $N = 10$ the same function (12) describes the FID until $t \approx 2t_*$.

It is important that the time $t_* \approx 120$ practically does not depend on the number of spins.

The FT of the four signals in Figure 3 are very close – they all have the same width and the same two frequencies peaks.

For 12 quantum spins, there are $2^{12} = 4096$ energy states, which is why the Fourier spectrum appears continuous. Rather than listing all frequencies for 12 spins, we illustrate this with the eigenvalues of Eq. (2) for 5 spins (Fig. 3(a)), rounded to four decimal places:

E = [-2.5030 -1.5106 -1.5056 -1.5009 -1.4951 -1.4911 -0.5172 -0.5138 -0.5095 -0.5044 -0.5000 -0.4990 -0.4954 -0.4933 -0.4850 -0.4768 0.4828 0.4862 0.4905 0.4957 0.5000 0.5011 0.5046 0.5068 0.5150 0.5233 1.4895 1.4945 1.4992 1.5050 1.5091 2.4972].

The non-degenerate energy $E \approx -2.5$ corresponds to the magnetic quantum number $m = 5/2$; similarly $E \approx 2.5$ corresponds to $m = -5/2$; states with other $m$ values are degenerate. The intervals between energy states are determined by the parameter $p_d \ll 1$, and for a substantial number of spins the energy spectrum looks quasi-continuous.

### 3.1 Classical spins equations

The dynamics of classical variables analogous to the quantum functions (10) and (11) can be derived either from the gyration equation for classical magnetic moments, or by taking the Heisenberg equation for operators and performing a classical mean-field substitution. The latter approach maintains a direct link to quantum formalism.

The Heisenberg equation for operator of the $l$-th spin magnetic moment $\hat{\vec{\mu}}_l$ in a field $\vec{H}$ is:

$$\dot{\hat{\vec{\mu}}}_l = -\frac{i}{\hbar}\left[\hat{\vec{\mu}}_l, \hat{\mathcal{H}}\right]. \tag{13}$$

The Hamiltonian is $\hat{\mathcal{H}} = -\sum_l \hat{\vec{\mu}}_l \vec{H}$, where the field $\vec{H} = \vec{H} + \vec{H}_l$, $\vec{H}$ is an external field and $\vec{H}_l$ is the dipole magnetic field at the location of $l$-th spin.

From equation (1a) we obtain (for details, see [6])

$$\dot{\hat{\mu}}_l^\delta = \gamma_l \sum_k e_{\delta\alpha\sigma} \left( \hat{\mu}_l^\alpha H^\sigma + D_{lk}^{\alpha\beta} \hat{\mu}_k^\beta \hat{\mu}_l^\sigma \right), \tag{14}$$

where $D_{lk}^{\alpha\beta} = \dfrac{1}{r_{lk}^3}\delta_{\alpha\beta} - \dfrac{3}{r_{lk}^5} r_{lk}^\alpha r_{lk}^\beta$, Greek letters denote vectors components, and Latin letters label spins, $H^\sigma$ is an external field (it can have all three component but in our calculations $\vec{H} = (0, 0, H_0)$).

Alternatively, the magnetic moments can be treated as classical vectors from the outset. For a magnetic moment $\vec{\mu}_l$ of the $l$-th particle, the gyration equation is

$$\frac{d}{dt}\vec{\mu}_l = |\gamma_l| \vec{\mu}_l \times \left( \vec{H}_0 + \vec{H}_l \right). \tag{15}$$

The dipole energy of a system of classical magnetic moments is

$$E_{dd} = \frac{1}{2} \sum_{l \neq k} D_{lk}^{\alpha\beta} \mu_l^\alpha \mu_k^\beta. \tag{16}$$

From this, the dipole field $\vec{H}_l$ at the location of the $l$-th particle is obtained as

$$H_l^\alpha = -\partial E_{dd} / \partial \mu_l^\alpha = -\frac{1}{2} \sum_k \left( D_{lk}^{\alpha\beta} \mu_k^\beta + D_{kl}^{\alpha\beta} \mu_k^\beta \right) = -\sum_k D_{lk}^{\alpha\beta} \mu_k^\beta. \tag{17}$$

Substituting (17) into (15) gives

$$\dot{\mu}_l^\delta = \gamma_l e_{\alpha\sigma\delta} \left( H^\alpha + H_l^\alpha \right) \mu_l^\sigma = \gamma_l \sum_k e_{\delta\alpha\sigma} \left( \mu_l^\alpha H^\sigma + D_{lk}^{\alpha\beta} \mu_k^\beta \mu_l^\sigma \right). \tag{18}$$

Equations (14) and (18) reveal the correspondence between the classical and quantum pictures of interacting magnetic moments: the Heisenberg equation for operators $\hat{\vec{\mu}}_l$ and the gyration equation (18) for classical vectors $\vec{\mu}_l$ lead to equations identical forms, even in the presence of spin-spin interaction [6] (this equivalence is trivial for non-interacting spins).

To relate equations (14) and (18), however, the operators in (14) must be replaced by their expectation values. In this way, the equations for $\langle \hat{\mu}_l^\sigma \rangle$ can be regarded as identical to the gyration equations for classical magnetic moments, $\mu_l^\sigma$, only if $\langle \hat{\mu}_k^\beta \hat{\mu}_l^\sigma \rangle$ is approximated as $\langle \hat{\mu}_k^\beta \rangle \langle \hat{\mu}_l^\sigma \rangle$. If this mean-field approximation is accepted, it provides a rational for the common practice of modeling spin dynamics with classical spins.

Nevertheless, the substitution of $\langle \hat{\mu}_i^\alpha \hat{\mu}_j^\beta \rangle$ with $\langle \hat{\mu}_i^\alpha \rangle \langle \hat{\mu}_j^\beta \rangle$ is questionable. Furthermore, while the Schrödinger equation is linear, the classical spins equations (18) are nonlinear. These differences naturally lead to discrepancies between the quantum and classical approaches.

To write equations (18) in the form more practical for simulations, we define angular frequencies associated with the constant field $H_0$, the average dipolar field $H_d = \mu / a^3$, and their ratio:

$$\omega_0 = |\gamma| H_0, \quad \omega_d = |\gamma| H_d, \quad p_d = \omega_d / \omega_0 = \frac{\mu}{a^3 H_0} = \frac{\gamma \mu}{a^3 \omega_0}, \tag{19}$$

where $\mu$ is the average magnetic moment magnitude (in the simulations below we assume particles with identical spins). This redefinition reduces the number of parameters so that the equations depend only on a single parameter, $p_d$. Note that $H_d$ (as well as $\omega_d$) represents the mean dipole field which is used only to define the parameter $p_d$, the actual local field at each spin is calculated using equation (17). Also note, that despite we use the same notation in (19) for the parameter $p_d$ as in case of quantum spins, the classical $p_d$ is not directly related with the spin value 1/2, one can only say that the value of $\mu$ in (19) is about Bohr's magneton $\mu_B$ and classical $p_d$ should be of the same order of magnitude as the quantum one. In simulations in this section we use the same value of $p_d = 0.01$ as for quantum spins in section 2, and obtain the value $t_*$ similar (but not the same) to the case of quantum spins. Note, that establishing a direct correspondence between the classical and quantum interaction constants is ambiguous. One can set the classical constant equal to the quantum one multiplied by $\sqrt{S(S+1)}$. Alternatively, one can obtain it from Hamiltonian (1), where the ratio of the dipole to Zeeman terms is proportional to the gyromagnetic ratio $\gamma$ that is twice as large for quantum spins as for classical spins. Crucially, our simulations demonstrate that regardless of the rescaling method chosen, the evolution of classical spins - which is highly sensitive to the details of the initial spin distribution (see Section 3.3.2) - matches the quantum evolution only in limited cases.

In simulations it is convenient to use dimensionless dipole field $\tilde{\vec{H}}_l$ and unit vectors of magnetic moments $\vec{e}_l$ (since equation (15) conserves the length of the magnetic moment) for the $l$-th spin, and a dimensionless time $\tilde{t}$:

$$\vec{H}_l / H_0 = p_d \tilde{\vec{H}}_l, \quad \vec{e}_l = \vec{\mu}_l / \mu, \quad \tilde{t} = \omega_0 t. \qquad (20)$$

Using these definitions, equations (18) become

$$\begin{cases} \dot{e}_k^x = e_k^y + p_d \left( e_k^y \tilde{H}_k^z - e_k^z \tilde{H}_k^y \right), \\ \dot{e}_k^y = -e_k^x + p_d \left( -e_k^x \tilde{H}_k^z + e_k^z \tilde{H}_k^x \right), \\ \dot{e}_k^z = p_d \left( e_k^x \tilde{H}_k^y - e_k^y \tilde{H}_k^x \right), \end{cases} \qquad (21)$$

where $\tilde{H}_k$ is the local dimensionless dipolar magnetic field at the $k$-th site, calculated as

$$\boldsymbol{H}_k / H_0 = p_d \tilde{\boldsymbol{H}}_k, \quad \tilde{\boldsymbol{H}}_k = \sum_{\substack{l=1 \\ l \neq k}}^{N} \left[ \frac{3}{\tilde{r}_{kl}^5} \tilde{\boldsymbol{r}}_{kl} \left( \boldsymbol{e}_l \tilde{\boldsymbol{r}}_{kl} \right) - \frac{1}{\tilde{r}_{kl}^3} \boldsymbol{e}_l \right], \qquad (22)$$

$\tilde{r}_{kl} = (r_l - r_k)/a$ are dimensionless vectors connecting the lattice sites.

Notice that for classical spins we use notation $e^x(t)$ to distinguish classical magnetic moments from quantum spins $S^x(t)$.

### 3.2 Algorithm of computations

The time evolution of FID is obtained by numerically integrating the system of $3N$ equations (21). We use adaptive algorithms, like Dormand-Prince method, Runge-Kutta and others with an automatically adjusted discretization time step with the precision of $10^{-8}$. Because the length of individual spin vectors must remain conserved, this constraint was exploited to control numerical accuracy: if after a single step the length of any unit spins deviated from 1 by more than $10^{-8}$, the step size was halved.

The Monte Carlo -based algorithm for preparing initial spin distributions with a specified polarization is described in [12].

The figures below show the transverse polarization evolution $e^x(t) = (1/N) \sum_l e_l^x(t)$ for a system of classical spins sitting in each node of a regular cubic lattice with the sides $N_x = N_y = N_z$ (except Figure 5b and Figure 6 for a linear chain). All spins are coupled via the dipolar interaction.

### 3.3. Comparison of results of quantum and classical simulations

The comparison of results of quantum and classical simulations reveals both similarities and significant discrepancies that are considered below. The origin of those discrepancies is the fundamental difference between the quantum dynamics governed by the linear Schrödinger equation (4) and the classical dynamics governed by the system of nonlinear equations (21).

### 3.3.1. Long-time behavior of solutions

The most drastic difference between the quantum and classical systems is the sensitivity of the classical dynamics to small variations of initial conditions that makes the long-time behavior of solutions unpredictable.

The solution of the Schrödinger equation (4) is characterized by a fixed finite set of frequencies corresponding to the eigenvalues of the Hamiltonian. It is constructed as a superposition of eigenfunctions with coefficients determined by the initial quantum state. Small changes in the initial conditions lead to a small change of the wavefunction and hence to small changes of averages for operators corresponding to physical variables. Thus, the evolution of a quantum system is not sensitive to small variations of initial conditions.

The classical system of interacting magnetic moments, governed by the nonlinear equations (21), is non-integrable. Depending on the initial conditions, its trajectories can be periodic, quasiperiodic, or chaotic. Since the system is integrable for $p_d = 0$, periodic and quasiperiodic trajectories prevail for small $p_d = 0$. As $p_d = 0$ increases, however, chaotic trajectories become dominant. These are characterized by an exponential divergence of initially close trajectories.

As an example, let us consider simulations with nearly 100% initial transverse polarization, as in the quantum case above. Figure 4 shows the FID signal for two runs with almost identical initial conditions: the total $e^x(0)$ is the same, with individual spin $e^x(t)$ deviations of the order of $10^{-4}$. The $e^x(t)$ curves are nearly identical for $t < t_* \approx 120$ (remind that the time of the first zero of the FID amplitude, $t_*$, can be associated with the phenomenological spin-spin relaxation time $T_2$). However, at $t > t_*$ both curves diverge irreversibly, which is the sign of chaos.

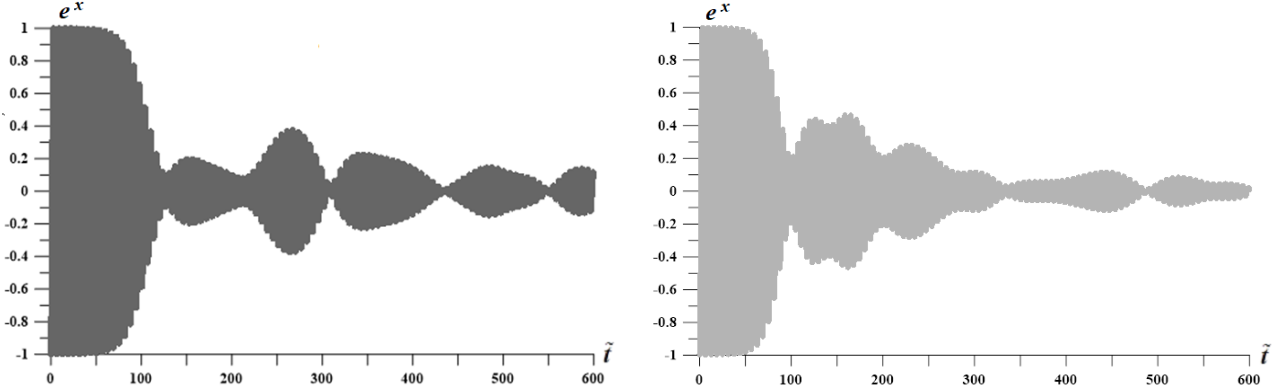

Fig. 4. Graph of $e^x(t)$. $N_{spins} = 5 \times 5 \times 5$, $p_d = 0.01$, $e^x(0) = 0.999999999$. Two panels represent two runs with identical parameters. The dimensionless time is $\tilde{t} = \omega_0 t$.

The irregular behavior of $e^x(t)$ at large $t$, which changes significantly due to tiny changes of initial conditions, is in sharp contrast to the regular and reproducible behavior of the quantum system.

To confirm the presence of chaos in the system of classical spins, we have calculated the maximum Lyapunov exponent [16]. For that goal, we have linearized nonlinear equations (21) around the solution $\vec{e}_i(t)$ and obtained the system of linear equation for disturbances $\vec{f}_i(t)$ presented in Appendix A. That system was integrated in parallel with the nonlinear system (21) itself.

The finite-time Lyapunov exponent was calculated as

$$L(t) = \frac{1}{t-T} \ln \frac{\|f(t)\|}{\|f(T)\|}, \qquad t > T, \qquad (23)$$

where

$$\|f(t)\| = \sqrt{\sum_{k=1}^{N} \left[ \left(f_k^x(t)\right)^2 + \left(f_k^y(t)\right)^2 + \left(f_k^z(t)\right)^2 \right]}$$

is the norm of the solution of the linear problem, $\vec{f}_i(t)$. In case of chaos functions $f_i^{x,y,z}$ grow exponentially, otherwise at large time it is slowly growing with time function. In (23), the initial fragment of the trajectory with $0 < t < T$ is not used by the computation of the Lyapunov exponent. The value of $T$ was chosen equal to 1000. The rate of the divergence of trajectories on that fragment corresponding to the specific choice of the initial state is not typical. The elimination of the initial fragment improves the convergence of $L(t)$ to its limit value,

$$\lim_{t \to \infty} L(t) = L_\infty,$$

that can be either equal to 0 or positive. In the latter case, the trajectory is chaotic.

Note that for the sufficiently precise calculation of $L_\infty$, we have to perform the integration during the time $t \sim 10^4$, which is significantly larger than $t_*$.

To analyze the chaos appearance, we calculated $L(t)$ for systems with different numbers of spins, different geometries and different values of the parameter of dipole interaction $p_d$. As an example, two typical situations with chaos are shown in Figure 5. Panel (a) represents the 3D case with $N_x = N_y = N_z = 5$, $p_d = 0.01$, $e^x(0) = 0.9999$ (the results do not depend on the initial polarization $f^x(0)$ in the linear problem, its particular value in this section was chosen $f^x(0) = 0.8$), panel (b) represents the 1D case with $N_x = 12$, $N_y = N_z = 1$, $p_d = 0.01$, $e^x(0) = 0.5$.

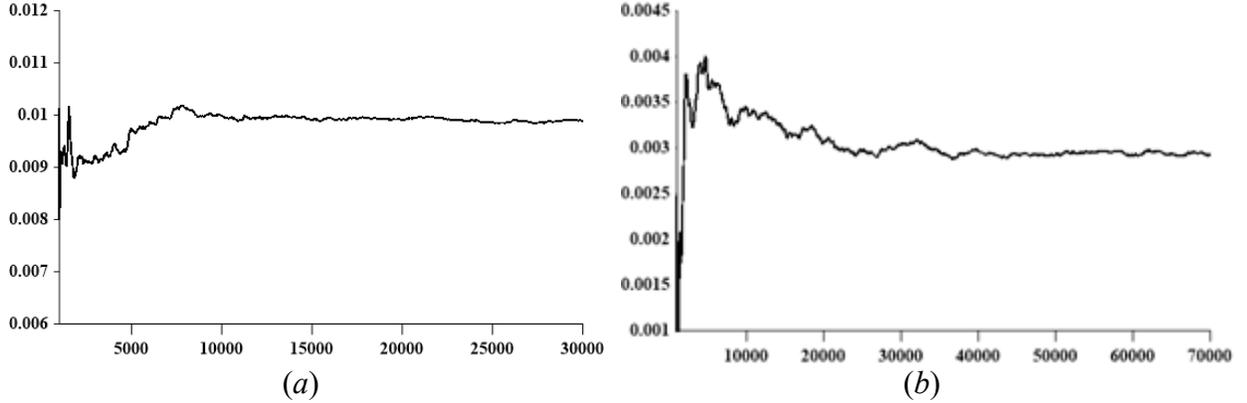

Fig. 5. Graph of $L(t)$. Panel a: $N_x = N_y = N_z = 5$, $e^x(0) = 0.9999$, $p_d = 0.01$. $L(30000) = 0.009889$.

Panel b: $N_x = 6$, $N_y = 1$, $N_z = 1$, $e^x(0) = 0.5$, $p_d = 0.01$. $L(70000) = 0.002927$

The same value $L(70000) = 0.002927$ and practically identical graph is obtained for 12 spins, $N_x = 12$, $N_y = 1$, $N_z = 1$.

For four spins, $N_x = 4$, $N_y = 1$, $N_z = 1$, functions $f_i^{x,y,z}$ still grow exponentially, but with smaller exponent than in case of six spins; the value $L(70000) = 0.001680$.

For comparison, we present a similar computation of $L_\infty$ in the case of two and three spins, where there is no chaos - Figure 6 shows that in this case $L(t) \to 0$.

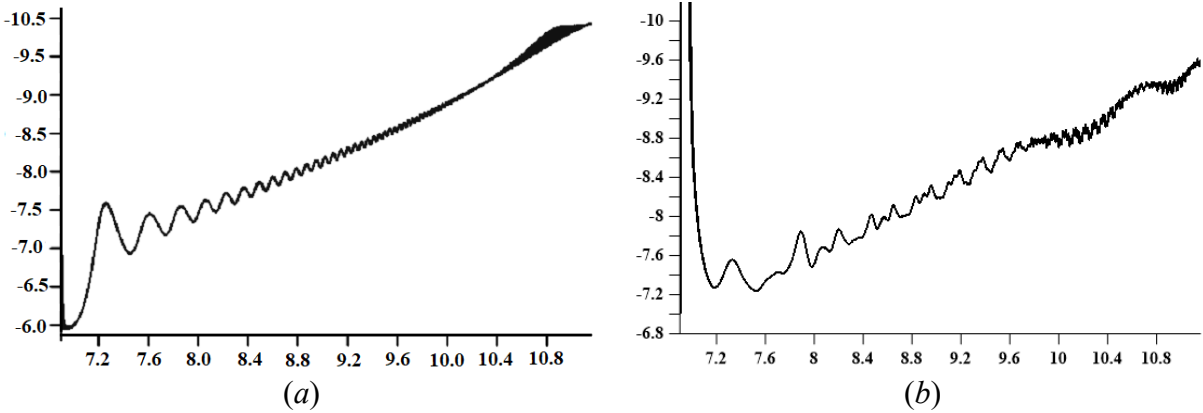

Fig. 6. $\ln(L(t))$ versus $\ln(t)$. $e^x(0) = 0.5$, $p_d = 0.01$.
Panel a: $N_x = 2$, $N_y = 1$, $N_z = 1$, $L(70000) = 0.0000587$.
Panel b: $N_x = 3$, $N_y = 1$, $N_z = 1$, $L(70000) = 0.00006837$.

We can conclude that for $N > 3$ spins the classical spins dynamics (for the value $p_d = 0.01$) are chaotic.

We also have analyzed the dependence of the Lyapunov exponent $L_\infty$ on the dipole interactions parameter $p_d$. For small value of $p_d$, such as $p_d = 10^{-4}$, we did not observe an exponential growth of functions $f_i^{x,y,z}$ - apparently, the classical trajectory corresponding to this value of $p_d$ was not chaotic. The values of $L_\infty$ calculated for $p_d = 0.005$, $p_d = 0.01$ and $p_d = 0.1$, are proportional to $p_d$ and match the observed time of the divergence of trajectories, $L_\infty \sim 1/t_*$.

A full analysis of chaotic and quasiperiodic components in a high-dimensional phase space is a formidable task, and such a detailed characterization was not the goal of this paper. For $p_d \gtrsim 0.005$ the runs with different initial conditions yielded practically identical values of the Lyapunov exponent. Therefore, we expect that the measure of the observed chaotic component is sufficiently high.

### 3.3.2. Short-time behavior of solutions

As mentioned above, the Fourier spectrum of any solution of the linear Schrödinger equation is a subset of the same finite set of eigenvalues of the Hamiltonian. For a classical system, the Fourier spectrum is different for each solution. Even in the simplest case of two interacting spins (see Appendix B), where all the trajectories are periodic or quasiperiodic, the set of frequencies is infinite, because of the presence of multiple and combinational frequencies, and it is different for different trajectories. Like in the case of two classical spins where the dynamics strongly depends on the initial conditions, similar situation is for many classical spins. Moreover, the evolution of the system depends on the type of the initial distribution – the "linear distribution", when all spins initially are directed only along or against the chosen axis, or the "random distribution", when spins can initially have any directions. Specifically, for the same overall value of the system polarization $e^x(0)$, the evolution for the time up to $t_*$ and even several $t_*$ for these two kinds of initial distribution is very different. For linear distribution, the short-time FID behavior is characterized by a long plateau before time $t_*$ (this behavior also does match the quantum spins FID). The length of this plateau for a given value of the parameter $p_d$ practically does not depend on the initial polarization for a wide range of $e^x(0)$, from high, like in Figure 4, to low, like $e^x(0) = 0.1$. Because of such a plateau, the classical spins FID in case of initial linear distribution cannot be satisfactory described by the phenomenological linear response MR function (12). Also note, that for initial polarization as low as $e^x(0) = 0.1$, the FID beats decrease very slowly, for instance, the FID second maximum is about 2/3 of the initial $e^x(0)$.

For a random distribution, there is no long plateau in the FID signal, and it is described by function (12), as in the quantum case.

To demonstrate these features, the results for FID simulations for $N = 7^3$ spins are shown in Figure 7.

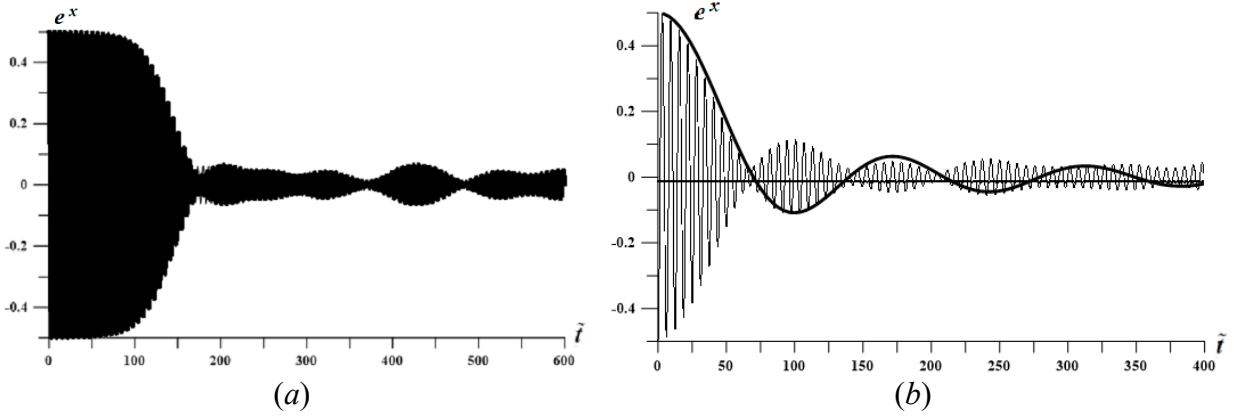

Fig. 7. $N = 7^3$, $p_d = 0.01$, $e^x(0) = 0.5$. Panel a: linear initial distribution, panel b: random initial distribution.

Solid line on panel b represents function (12) with the parameters $A = 0.5$, $a = 0.0009$, $b = 0.045$.

Very close results are obtained for $N = 5^3$ spins and $e^x(0) = 0.5$ with the same, as in Figure 7, value $t_* \approx 170$ for a linear distribution, and $t_* \approx 70$ up to the end of the second beat for a random distribution. The value of $t_*$ depends only on $p_d$ and remains stable across different statistical ensembles for a selected polarization type, and is largely independent of $e^x(0)$.

The difference between random and linear distributions demonstrates, in another form, the system's sensitivity to initial conditions. Thus, for classical spins, specific details are crucial – the dynamics depend strongly on the initial distribution. An important observation is that the position of the first zero of $e^x(t)$, which determines the duration of the FID signal, depends on the method used to choose the initial polarization (for very high polarization, like in Fig. 4, the results for linear and random distributions are the same). Intuitively, a linear distribution seems closer to the quantum case, where initial spins are aligned strictly along or against the assigned axis. However, more physically realistic FID results with classical spins are obtained with a random distribution.

### 3.3.3. Dependence on the number of spins

How well do classical spin simulations describe macroscopic behavior like FID when the number of spins is limited? In Figures 8 and 9 we consider different number of spins, $N = 7^3$, $N = 5^3$ and $N = 3^3$.

For several tens of spins, collective decay already appears. The early-stage decay for $N = 7^3$ and $N = 5^3$ looks similar, but to reproduce the long-time decreasing of oscillation amplitudes at $t \gg 1/\omega_d$, several hundred spins are required. One can say that when the number of classical spins exceeds about a hundred, it provides an acceptable qualitative description, but still the FID signal is not vanishing.

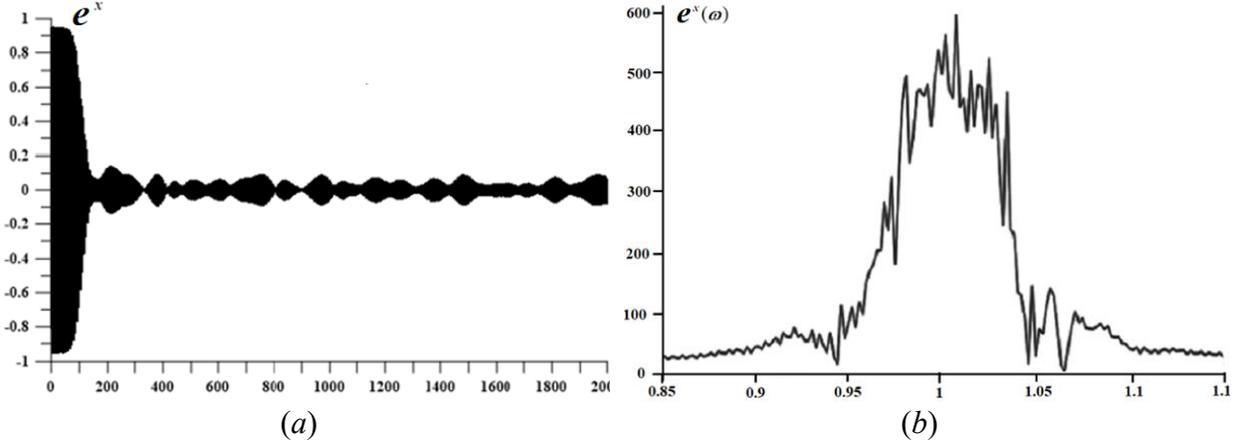

Fig. 8. $N = 7 \times 7 \times 7$, $p_d = 0.01$, $e^x(0) = 0.95$.

Panel a: Graph $e^x(t)$. Panel b. Fourier transform of $e^x(t)$.

The Fourier transform in Figure 8b has the same spectral width as that of the quantum spins in Figure 1. However, the fine structure - such as the two distinct "Pike's Peak" features – is not reliably resolved. Recall that while the spectrum of a quantum spin system is discrete, the classical chaotic dynamics is characterized by the continuous spectrum. That continuous spectrum, which is produced by the late stages of the classical system evolution, obscures the peaks that correspond to the early stage of evolution. The Fourier analysis carried out for a shorter initial time interval gives a frequency spectrum more similar to the quantum calculations.

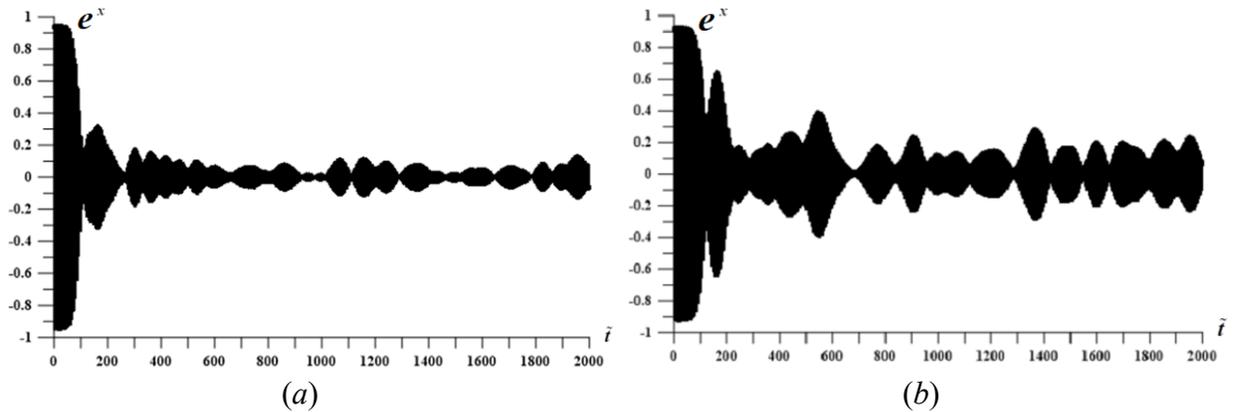

Fig. 9. Graph $e^x(t)$, $p_d = 0.01$, $e^x(0) = 0.95$. Panel a: $N = 5 \times 5 \times 5$. Panel b: $N = 3 \times 3 \times 3$.

Thus, modeling the time evolution of a classical spin system requires a much larger number of spins than its quantum counterpart.

**Conclusion**

One of the goals of this paper is to explore the limits of classical spins calculations for macroscopic quantities such as FID. The classical spins approach is often used to describe various spin phenomena with the commonly held assumption that for a large number of spins it can adequately approximate the macroscopic behavior of quantum systems. In fact, for large spin systems classical spin simulations give a reasonable overall description of FID. However, as we have shown, significant differences arise between quantum and classical cases, particularly at the beginning of evolution and on long timescales.

A fundamental difference exists between quantum and classical spin dynamics. The quantum problem is linear; therefore, the temporal spectrum of oscillations of any matrix element is finite (though possibly large) and discrete, consisting solely of differences between Hamiltonian eigenvalues. In contrast, the classical problem is nonlinear, and the spectrum of any observable quantity is typically either discrete with an infinite number of frequencies (quasiperiodic motion) or continuous (chaotic motion), depending on the initial conditions.

We performed numerical simulations of FID in a system of dipole-interacting classical spins and provide the direct solution of the Schrödinger equation in the case of quantum spins. The overall FID patterns show qualitative similarities, but a detailed comparison reveals significant differences in both short and long timescales.

For times shorter than the characteristic dipole-dipole interaction time, the quantum and classical systems having the same initial polarization demonstrate very different dynamics: the quantum system exhibits an exponential FID decay from the very beginning of evolution, whereas the classical system demonstrates a long, nearly steady evolution before a rapid decay sets in.

Furthermore, we found that to achieve an adequate description of time evolution in a classical spin system both on short and long timescales, a much larger number of spins is required compared to the corresponding quantum system.

Studying the dynamics of a classical system requires solving a large system of differential equations – a task that comes with all the associated numerical difficulties. Our quantum approach, by contrast, does not involve integrating equations of motion. Instead, it is based entirely on three steps: (i) diagonalizing the Hamiltonian to obtain its spectrum, (ii) constructing the time-dependent wave function of the system, and (iii) computing expectation values.

To analyze the discrepancy at times longer than the characteristic dipole time, we examined chaotic behavior in a system of classical spins and have found that in a system of more than three classical spins the dynamics becomes chaotic for the value of the dipole interaction parameter (the ratio of dipole to Zeeman energies) $p_d \gtrsim 0.005$.

## Appendix A. Classical equations

Equations (21) for unit vectors $\vec{e}_i = \vec{\mu}_i / \mu$, $i = 1,...,N$ can be written in the more explicit form which was used in simulations:

$$\dot{e}_i^x = e_i^y + p_d \sum_{j \neq i} \left[ -a_{ij}\left(e_i^y e_j^z + \frac{1}{2} e_i^z e_j^y\right) \right.$$
$$\left. - 2c_{ij}^x e_i^y e_j^x + 2c_{ij}^y \left(e_i^y e_j^y - e_i^z e_j^z\right) - 2e_i^z \left(b_{ij}^x e_j^y + b_{ij}^y e_j^x\right) \right],$$

$$\dot{e}_i^y = -e_i^x + p_d \sum_{j \neq i} \left[ a_{ij}\left(e_i^x e_j^z + \frac{1}{2} e_i^z e_j^x\right) \right. \quad\quad\quad (A1)$$
$$\left. + 2c_{ij}^x \left(e_i^x e_j^x - e_i^z e_j^z\right) - 2c_{ij}^y e_i^x e_j^y - 2e_z^{(i)}\left(b_{ij}^x e_j^x - b_{ij}^y e_j^y\right) \right],$$

$$\dot{e}_i^z = p_d \sum_{j(\neq i)} \left[ \frac{1}{2} a_{ij}\left(e_i^x e_j^y - e_i^y e_j^x\right) \right.$$
$$\left. + 2e_z^{(j)}\left(c_{ij}^x e_i^y + c_{ij}^y e_i^y\right) + 2b_{ij}^x\left(e_i^y e_j^y + e_i^x e_j^x\right) + 2b_{ij}^y\left(e_i^y e_j^x - e_i^x e_j^y\right) \right].$$

The coefficients in these equations originate from the secular $\hat{\mathcal{H}}_d^s$ and non-secular $\hat{\mathcal{H}}_d^{ns}$ parts of the dipole-dipole Hamiltonian:

$$\hat{\mathcal{H}}_d^s = (\gamma\hbar)^2 \sum_{l>k} a_{lk}\left[\hat{S}_l^z \hat{S}_k^z - \frac{1}{4}(\hat{S}_l^+ \hat{S}_k^- + \hat{S}_l^- \hat{S}_k^+)\right], \quad\quad (A2)$$

$$\hat{\mathcal{H}}_d^{ns} = (\gamma\hbar)^2 \sum_{l>k}\left[b_{lk}\hat{S}_l^+ \hat{S}_k^+ + b_{lk}^*\hat{S}_l^- \hat{S}_k^- + 2c_{lk}\hat{S}_l^+ \hat{S}_k^z + 2c_{lk}^*\hat{S}_l^- \hat{S}_k^z\right]. \quad\quad (A3)$$

Here

$$a_{ij} = \frac{1}{r_{ij}^3}\left[1-3\frac{(z_i-z_j)^2}{r_{ij}^2}\right], \quad b_{ij}^x = -\frac{3}{4r_{ij}^5}\left[(x_i-x_j)^2-(y_i-y_j)^2\right], \quad b_{ij}^y = \frac{3}{2r_{ij}^5}(x_i-x_j)(y_i-y_j),$$

$$c_{ij}^x = -\frac{3}{2r_{ij}^5}(x_i-x_j)(z_i-z_j), \quad c_{ij}^y = \frac{3}{2r_{ij}^5}(y_i-y_j)(z_i-z_j);$$ vectors $r_{ij} = (x_{ij}, y_{ij}, z_{ij})$ are in units of the lattice step.

In contrast to $\hat{\mathcal{H}}_d^s$, $\hat{\mathcal{H}}_d^{ns}$ does not commute with the z-component of the total magnetic moment. If the non-secular part is neglected, transforming to a rotating frame with frequency $\omega_0 = |\gamma| H_0$ eliminates the Zeeman term in the Hamiltonian (1) and allows calculations with a substantially bigger time step. In many of our calculations, however, we use the laboratory frame, because the primary limitation is memory (RAM) rather than computation speed. Moreover, working in the laboratory frame enables us to examine the role of the non-secular Hamiltonian.

*Linearized classical spins equations*

To obtain the linearized equations, we must replace terms $e_i^{x,y,z}$ in the classical spins system with $e_i^{x,y,z} + f_i^{x,y,z}$ and collect all terms of the first order in $f$. That means that in the linear terms in equations (A1) we replace $e_i^{x,y,z} \to f_i^{x,y,z}$, while the quadratic terms are changed in the way, for instance $e_i^y e_j^z \to e_i^y f_j^z + e_j^z f_i^y$, etc. It gives the following equations for $f_i^{x,y,z}$:

$$\dot{f}_i^x = f_i^y + p_d \sum_{j \neq i}\left[-a_{ij}\left(e_i^y f_j^z + e_j^z f_i^y + \frac{1}{2}(e_i^z f_j^y + e_j^y f_i^z)\right)\right.$$
$$\left. -2c_{ij}^x\left(e_i^y f_j^x + e_j^x f_i^y\right) + 2c_{ij}^y\left(e_i^y f_j^y + e_j^y f_i^y - e_i^z f_j^z - e_j^z f_i^z\right)\right. \quad (A4)$$
$$\left. -2e_i^z(b_{ij}^x f_j^y + b_{ij}^y f_j^x) - 2(b_{ij}^x e_j^y + b_{ij}^y e_j^x) f_i^z\right],$$

$$\dot{f}_i^y = -f_i^x + p_d \sum_{j \neq i} \left[ a_{ij} \left( e_i^x f_j^z + e_j^z f_i^x + \frac{1}{2} \left( e_i^z f_j^x + e_j^x f_i^z \right) \right) \right.$$
$$+ 2c_{ij}^x \left( e_i^x f_j^x + e_j^x f_i^x - e_i^z f_j^z - e_j^z f_i^z \right) - 2c_{ij}^y \left( e_i^x f_j^y + e_j^y f_i^x \right) \tag{A5}$$
$$\left. -2e_i^z \left( b_{ij}^x f_j^x - b_{ij}^y f_j^y \right) - 2 \left( b_{ij}^x e_j^x - b_{ij}^y e_j^y \right) f_i^z \right],$$

$$\dot{f}_i^z = p_d \sum_{j(\neq i)} \left[ \frac{1}{2} a_{ij} \left( e_i^x f_j^y + e_j^y f_i^x - e_i^y f_j^x - e_j^x f_i^y \right) + 2e_j^z \left( c_{ij}^x f_i^y + c_{ij}^y f_i^x \right) \right.$$
$$+ 2 \left( c_{ij}^x e_i^y + c_{ij}^y e_i^x \right) f_j^z + 2b_{ij}^x \left( e_i^x f_j^y + e_j^y f_i^x + e_i^y f_j^x + e_j^x f_i^y \right) \tag{A6}$$
$$\left. + 2b_{ij}^y \left( e_i^x f_j^x + e_j^x f_i^x - e_i^y f_j^y - e_j^y f_i^y \right) \right].$$

The system of equations for functions $f_i^{x,y,z}$ is integrated along with the system (A1) for functions $e_i^{x,y,z}$ to calculate the Lyapunov exponent (23).

**Appendix B. The system of two spins**

**Two quantum spins**

First, consider the situation when two quantum ½ spins are placed on the z - axis – in this case the dipole Hamiltonian has only a secular part and the eigenfunctions and the eigenvalues $\varepsilon_i$ (in units of $\hbar \omega_0$) of the Hamiltonian (1) have the simplest form:

$$v_1 = \begin{pmatrix} 1 \\ 0 \\ 0 \\ 0 \end{pmatrix} \text{ for } \varepsilon_1 = -1 - p_d, \quad v_2 = \frac{1}{\sqrt{2}} \begin{pmatrix} 0 \\ -1 \\ 1 \\ 0 \end{pmatrix} \text{ for } \varepsilon_2 = 2p_d, \quad v_3 = \frac{1}{\sqrt{2}} \begin{pmatrix} 0 \\ 1 \\ 1 \\ 0 \end{pmatrix} \text{ for } \varepsilon_3 = 0, \quad v_4 = \begin{pmatrix} 0 \\ 0 \\ 0 \\ 1 \end{pmatrix}$$

for $\varepsilon_4 = 1 - p_d$. \hfill (B1)

In this case the wavefunction satisfying the initial condition $\hat{S}^x \Psi_0 = \frac{N}{2} \Psi_0$ with $\hat{S}^x = \hat{S}_1^x + \hat{S}_2^x$, $N = 2$, is

$$\Psi(t) = \frac{1}{2} \left( e^{-i(1+p_d)t}, 1, 1, e^{i(1-p_d)t} \right)^T. \tag{B2}$$

It gives

$$S^x(t) = \Psi^*(t)\hat{S}^x\Psi(t) = \cos t \cos(p_d t). \tag{B3}$$

**Two classical spins**

Let us consider now the system of two classical magnetic moments located on the axis $z$ on the distance $a$ of each other. To simplify the notations, we define
$$x_i \equiv e_i^x, \quad y_i \equiv e_i^y, \quad z_i \equiv e_i^z.$$
Then system $z$ is reduced to
$$\frac{dx_1}{dt} = y_1 + p(2y_1 z_2 + y_2 z_1), \quad \frac{dy_1}{dt} = -x_1 - p(2x_1 z_2 + x_2 z_1), \quad \frac{dz_1}{dt} = p(-x_1 y_2 + y_1 x_2), \tag{B4}$$
$$\frac{dx_2}{dt} = y_2 + p(2y_2 z_1 + y_1 z_2), \quad \frac{dy_2}{dt} = -x_2 - p(2x_2 z_1 + x_1 z_2), \quad \frac{dz_2}{dt} = p(-x_2 y_1 + y_2 x_1), \tag{B5}$$
where $p \equiv p_d$ is the dipole-dipole interactions parameter (6b). In the rotating system of coordinates,
$$x_i = X_i \cos t + Y_i \sin t, \quad y_i = Y_i \cos t - X_i \sin t,$$
we obtain:
$$\frac{dX_1}{dt} = p(2Y_1 z_2 + Y_2 z_1), \quad \frac{dY_1}{dt} = -p(2X_1 z_2 + X_2 z_1), \quad \frac{dz_1}{dt} = p(-X_1 Y_2 + Y_1 X_2), \tag{B6}$$
$$\frac{dX_2}{dt} = p(2Y_2 z_1 + Y_1 z_2), \quad \frac{dY_2}{dt} = -p(2X_2 z_1 + X_1 z_2), \quad \frac{dz_2}{dt} = p(-X_2 Y_1 + Y_2 X_1). \tag{B7}$$

Because the lengths of vectors are conserved, $X_1^2 + Y_1^2 + z_1^2 = 1$ and $X_2^2 + Y_2^2 + z_2^2 = 1$, it is reasonable to use the spherical system of coordinates,
$$X_1 = \sin\theta_1 \cos\varphi_1, \quad Y_1 = \sin\theta_1 \sin\varphi_1, \quad z_1 = \cos\theta_1,$$
$$X_2 = \sin\theta_2 \cos\varphi_2, \quad Y_2 = \sin\theta_2 \sin\varphi_2, \quad z_2 = \cos\theta_2.$$
Substituting these expressions into equations for $X_i$, $Y_i$, $z_i$, $i = 1, 2$, and taking the corresponding linear combinations of equations for $X_i$ and $Y_i$, we obtain the following system of equations for angles:
$$\dot\theta_1 = -p\sin\theta_2 \sin(\varphi_1 - \varphi_2), \quad \dot\theta_2 = p\sin\theta_1 \sin(\varphi_1 - \varphi_2),$$
$$\dot\varphi_1 = -1 + p[-2\cos\theta_2 - \sin\theta_2 \cot\theta_1 \cos(\varphi_1 - \varphi_2)],$$
$$\dot\varphi_2 = -1 + p[-2\cos\theta_1 - \sin\theta_1 \cot\theta_2 \cos(\varphi_1 - \varphi_2)].$$
Obviously, the dynamics is determined by the difference of angles $\varphi_1 - \varphi_2$ rather than $\varphi_1$ and $\varphi_2$ (because the system of two spins on one $z$-axis is rotationally invariant). Therefore, we can introduce $\Phi = \varphi_1 - \varphi_2$ and rewrite the equations using variables $\theta_1$, $\theta_2$ and $\Phi$. The coefficient $p$ can be scaled out by the change of the time variables, $d/dt = pd/d\tau$. Finally, we obtain the following universal equations for the evolution of slowly evolving variables:

$$\frac{d\theta_1}{d\tau} = -\sin\theta_2 \sin\Phi, \tag{B8}$$

$$\frac{d\theta_2}{d\tau} = \sin\theta_1 \sin\Phi, \tag{B9}$$

$$\frac{d\Phi}{d\tau} = 2(\cos\theta_1 - \cos\theta_2) + (\sin\theta_1 \cot\theta_2 - \sin\theta_2 \cot\theta_1)\cos\Phi. \tag{B10}$$

From (B8) and (B9), one can see that
$$M = \cos\theta_1 + \cos\theta_2 \tag{B11}$$
is the integral of motion (z-component of the moment). That means that the Zeeman energy is conserved. Another integral of motion is the dipole interaction energy which is equal to $p$ multiplied by
$$E = \sin\theta_1 \sin\theta_2 \cos\Phi - 2\cos\theta_1 \cos\theta_2. \tag{B12}$$

Let us denote $c = \cos\theta_1$. Taking into account that
$$\cos\theta_2 = M - \cos\theta_1 = M - c$$
and
$$\cos\Phi = \frac{E + 2\cos\theta_1 \cos\theta_2}{\sin\theta_1 \sin\theta_2} = \pm\frac{E + 2c(M-c)}{\sqrt{(1-c^2)[1-(M-c)^2]}},$$
we find the closed equation for $c(\tau)$:

$$\frac{dc}{d\tau} = \pm\sqrt{(1-c^2)[1-(M-c)^2] - [E + 2c(M-c)]^2}, \tag{B13}$$

which can be solved for any $M$ in the interval $-2 \le M \le 2$.

For our goal, it is sufficient to consider the particular case $M = 0$ corresponding to the initial conditions $\mathbf{e}_1 + \mathbf{e}_2 = s\mathbf{e}_x$. In that case $\cos\theta_1 = c$, $\cos\theta_2 = -c$, $\sin\theta_1 = \sqrt{1-c^2}$, $\cos\Phi = (E - 2c^2)/(1-c^2)$, and equation (B13) becomes

$$\frac{dc}{d\tau} = \pm\sqrt{(1 + E - 3c^2)(1 - E + c^2)}. \tag{B14}$$

The expression under the root is non-negative if (i) $1 + E - 3c^2 \ge 0$, $1 - E + c^2 \ge 0$ or (ii) $1 + E - 3c^2 < 0$, $1 - E + c^2 \le 0$.

In the case (ii), $(1+E)/3 < c^2 \le E - 1$, hence $E > 2$ and $c^2 > 1$, which is not possible.

In the case (i), we find that $E \le 2$. Equation (B14) has constant solutions, $c_{min}^2 = E - 1$ and $c_{max}^2 = (1+E)/3$, and time-periodic solutions. The time-periodic solutions oscillate between $c_{min}^2 = E-1$ and $c_{max}^2 = (1+E)/3$. Because $c_{min}^2$, this kind of solution is possible only for $E \geq 1$. Note that while $c = c_{min}$, $\Phi = 0$, and while $c = c_{max}$, $\Phi = \pi$. The solution of (B14) is

$$c^2(\tau) = \frac{1+E}{3} - \frac{2(2-E)}{3} sn^2\left(\sqrt{1+E}(\tau-\tau_0)\right), \quad \tau_0 = const, \quad (B15)$$

where $sn$ is the elliptic sine with modulus

$$k^2 = \frac{2(2-E)}{E+1}. \quad (B16)$$

The temporal period of the solution (B15), which is equal to

$$T(E) = \frac{2}{E+1} K\left(\frac{2(2-E)}{E+1}\right), \quad (B17)$$

depends on $E$, and it is changed from $\pi/\sqrt{3}$ at $E = 2$ to infinity when $E \to 1$.

The initial conditions with direction of both magnetic moments close to the direction of the axis $x$, i.e., $\theta_1$ and $\theta_2$ close to $\pi/2$, and $\Phi$ close to 0, according to (B12), correspond to $E$ close to 1. In that case, $k^2$ is close to 1, therefore the oscillation period is long, in agreement with the numerical calculations shown in Figure 10.

In the special case $E = 1, M = 0$, equation (B14) has three kinds of solutions, a stationary solution $c_1 = 0$ and aperiodic solutions $c_{2,3}(\tau) = \sqrt{2/3} \cosh^{-1}(\sqrt{2}(\tau-\tau_0))$. The stationary solution $c_1 = 0$ corresponds to initial conditions $\theta_1 = \theta_2 = \pi/2$, $\Phi = 0$, i.e., $x_1(0) = x_2(0) = 1$. Solutions $c_{2,3}$ describe a homoclinic trajectory: under an arbitrary small initial disturbance, the system leave the vicinity of the point $c = 0$ and then returns to it as $t \to \infty$. Note that both kinds of behaviors are strongly different from the periodic behavior predicted in the case of a quantum system.

The opposite limit, $E = 2-\varepsilon$, $0 < \varepsilon \leq 1$, corresponds to the case where $\theta_1 \approx 0$ and $\theta_2 \approx \pi$, i.e., the direction of one spin is close to that of axis $z$, while the other spin is directed nearly in opposite direction. In that limit, the modulus of the elliptic function is small, and the elliptic sine is close to the trigonometric sine. The solution (B15) gives

$$c^2(\tau) \approx 1 - \frac{\varepsilon}{3}[2 - \cos(2\sqrt{3}(\tau-\tau_0))]. \quad (B18)$$

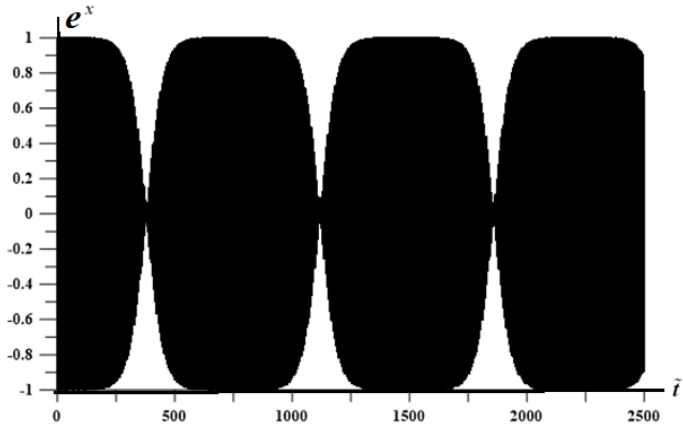

Fig. 10. $N_x = 1$, $N_y = 1$, $N_z = 2$, $p_d = 0.01$, $e^x(0) = 0.9999$